\documentclass[11pt]{article}

\usepackage[hmargin=2.8cm,vmargin=2.8cm]{geometry}
\usepackage{graphicx}
\begin{document}
\title{Generalized uncertainty relations and entanglement dynamics in quantum Brownian motion models}
\author{C. Anastopoulos\footnote{anastop@physics.upatras.gr}, S. Kechribaris\footnote{spiroskech@upatras.gr}, and D. Mylonas\footnote{dmylonas@upatras.gr}\\
Department of Physics, University of Patras, 26500 Patras, Greece}
\maketitle

\begin{abstract}
We study entanglement dynamics in quantum Brownian motion (QBM) models. Our main tool is the Wigner function propagator. Time evolution in the Wigner picture is physically intuitive and it leads to a simple derivation of a master equation  for any number of system harmonic oscillators and spectral density of the environment. It also provides generalized uncertainty relations, valid for any initial state, that allow a characterization of the environment in terms of the modifications it causes to the system's dynamics. In particular, the uncertainty relations are very informative about the entanglement dynamics of Gaussian states, and to a lesser extent for other families of states. For concreteness, we apply these techniques to a bipartite QBM model, describing the processes of entanglement creation, disentanglement, and decoherence at all temperatures and time scales.

\end{abstract}

\section{Introduction}

The study of quantum entanglement is both of practical and
theoretical significance: entanglement is viewed as a physical resource for
quantum-information processing  and it constitutes
a major  issue  in the foundations of quantum
theory. The quantification of entanglement is  difficult in multipartite systems (see, for example, Refs. \cite{PeresBook,KarolBook,AlickiBook, 4Hor}); however,  there are useful separability criteria and entanglement measures for bipartite states, pure and mixed \cite{Peres,  Horod, Simon, Duan, Barnum,  HHH, AlickiHorod, GMVT03}).

Realistic quantum systems, including multipartite ones,  cannot avoid interactions
with their environments, which can degrade their quantum coherence and
entanglement. Thus quantum decoherence and
disentanglement are obstacles to quantum-information processing \cite{RajRendell,Diosi,Dodd,DoddHal}. On the other hand, some environments act as intermediates that generate entanglement in multipartite systems, even if the components do not interact directly \cite{Braun, BFP, OK}.
The theoretical study of entanglement dynamics in open quantum systems has uncovered important physical effects, such as the sudden death of entanglement \cite{YE, suddeath}, entanglement revival after sudden death \cite{FicTan06}, the significance of non-Markovian effects \cite{ASH, nonmark, CYH}, the possibility of a rich phase structure for the asymptotic behavior of entanglement \cite{PR1, PR2}, and intricacies in the evolution of entanglement in multipartite systems \cite{Li10}.

Here, we study  entanglement and decoherence in quantum Brownian motion (QBM) models \cite{HPZ, QBM, QBM2}, focusing on their description in terms of generalized uncertainty relations. Our main  tool in this study is the Wigner function propagator. QBM models are defined by a quadratic total Hamiltonian, and they  are characterized by a Gaussian propagator.  This propagator is solely determined by two matrices: one corresponding to the classical dissipative equations of motion and one containing the effect of environment-induced diffusion. In Sec. II we provide explicit formulas for their determination.

The simplicity of time evolution in the Wigner picture leads to a concise derivation of an exact master equation  for general QBM models, with any number of system oscillators and spectral density.
Moreover, time evolution in the Wigner picture
allows for a derivation of generalized uncertainty relations, valid for {\em any} initial state, that incorporate the influence of the environment upon the system.  These uncertainty relations generalize the ones of Ref. \cite{AnHa} to QBM models with  an arbitrary number of system oscillators---see also Refs. \cite{HZ, CYH}. Their most important feature is that the lower bound is independent of the initial state, and for this reason, they allow for general statements about the process of decoherence and thermalization.

The uncertainty relations are also related to
 separability criteria for bipartite systems \cite{Simon, GMVT03}. Hence, they provide an important tool for the study of entanglement dynamics.
  For Gaussian states, in particular, the uncertainty relations, derived here, provide a general characterization of processes such as entanglement creation and disentanglement without the need to specify detailed properties of the initial state.  However,  uncertainty relations  do not suffice to distinguish all entangled {\em non-Gaussian} states. For such states, the description of entanglement dynamics from the uncertainty relations is rather partial, but still leads to nontrivial results.

  The uncertainty relations derived in this article apply to
   any open quantum system characterized by Gaussian propagation, and they are expressed solely in terms of the coefficients of the Wigner function propagator. They can be used for the study of entanglement dynamics, not only in bipartite but also in multipartite systems.
   To demonstrate their usefulness, we apply them to a concrete bipartite QBM model system that has been studied  by Paz and Roncanglia  \cite{PR1, PR2}. In this model, there exist two coupled subalgebras of observables, only one of which couples directly to the environment. For a special case of the system parameters, considered in Ref. \cite{PR1}, one of the subalgebras is completely decoupled, and thus there exists a decoherence-free subspace for the system. Here we focus on the generic case, also explored in Ref. \cite{PR2}.

    We find that in the high-temperature regime, decoherence and disentanglement are generic and the uncertainty relations allow for an identification of the characteristic timescales, which in some cases may be of very different orders of magnitude. At low temperature, entanglement creation often occurs and we demonstrate that it is  accompanied by ``entanglement oscillations'', that is, a sequence of entanglement sudden death and revivals at early times. In this regime, there is no decoherence, and disentanglement arises because of relaxation. At a time scale of the order of relaxation time the system tends to a unique asymptotic state, which coincides with a thermal state at the weak-coupling limit. The generalized uncertainty relations allow for the determination of upper limits to disentanglement time with respect to all Gaussian initial states.

The structure of the article is the following. In Sec. II we construct the Wigner function propagator for the most general QBM model and we provide explicit formulas for the propagator's coefficients. The master equation is then simply obtained from the propagator. In Sec. III we construct the generalized uncertainty relations valid for all QBM models, we show that they can be used for the study of multipartite entanglement,
and we then consider their special case in the model of Refs. \cite{PR1, PR2}. In Sec. IV we employ the uncertainty relations for the study of decoherence, disentanglement, and entanglement creation in different regimes and time scales of this model.


\section{Quantum Brownian motion models for multipartite systems}
In this section, we consider the most general setup for quantum Brownian motion, namely, a system of  $N$ harmonic oscillators of masses $M_r$ and frequencies $\Omega_r$  interacting with a heat bath. The heat bath is modeled  by a set of harmonic oscillators of masses $m_i$ and frequencies $\omega_i$, initially at a thermal state of temperature $T$. The Hamiltonian of the total system is a sum of three terms $\hat{H} = \hat{H}_{sys} + \hat{H}_{env} + \hat{H}_{int}$, where
\begin{eqnarray}
\hat{H}_{sys} &=& \sum_r\left( \frac{1}{2M_r}  \hat{P}_r^2 + \frac{M_r \Omega^2_r}{2} \hat{X}_r^2\right) \label{ho}\\
\hat{H}_{env} &=& \sum_i (\frac{1}{2m_i} \hat{p}_i^2 + \frac{m_i \omega_i^2}{2} \hat{q}_i^2)\\
\hat{H}_{int} &=& \sum_i \sum_a c_{ir} \hat{X}_r \hat{q}_i, \label{hint}
\end{eqnarray}
where $\hat{X}_r$ and $\hat{P}_r$ are the position and momentum operators for the system oscillators and $\hat{q}_i$ and $\hat{p}_i$ are the position and momentum operators for the environment oscillator. The interaction Hamiltonian Eq. (\ref{hint}) involves different couplings $c_{ir}$ of each system oscillators to the bath. Thus it can also be used to describes systems different from the classic setup of Brownian motion, for example, particle detectors at different locations interacting with a quantum field \cite{LinHu}.

For an initial state that is factorized in system and environment degrees of freedom the evolution of the reduced density matrix for the system variables is {\em autonomous}, and it can be expressed in terms of a master equation. For the issues we explore in this article, in particular entanglement dynamics, the determination of the propagator of the reduced density matrix is more important than the construction of the master equation, because it allows us to follow the time evolution of the relevant observables. The construction of the propagator is simpler in the Wigner picture.

\medskip

Instead of the density operator, we work with the Wigner function, defined by
\begin{eqnarray}
W({\bf X},{\bf P}) = \frac{1}{(2 \pi)^N} \int d^N \zeta e^{-i {\bf P} \cdot {\bf \zeta}} \hat{\rho}({\bf X} + \frac{1}{2}{\bf \zeta}, {\bf X}- \frac{1}{2}{\bf \zeta}).
\end{eqnarray}
Its inverse is
\begin{eqnarray}
\hat{\rho}({\bf X},{\bf Y}) = \int d^NP \; e^{i{\bf P} \cdot ({\bf X} - {\bf X'})} \; W(\frac{1}{2}({\bf X} + {\bf X'}), {\bf P}).
\end{eqnarray}

For a factorized initial state,  time evolution in QBM models is encoded in the density matrix propagator $J({\bf X}_f, {\bf Y}_f, t| {\bf X}_0, {\bf Y}_0,0)$, defined  by
\begin{eqnarray}
\hat{\rho}_t({\bf X}_f, {\bf Y}_f) = \int d^NX_0 d^NY_0 \; \; J({\bf X}_f, {\bf Y}_f, t| {\bf X}_0, {\bf Y}_0, 0) \hat{\rho}_0({\bf X}_0, {\bf Y}_0). \label{jprop}
\end{eqnarray}
The Wigner function propagator is defined as
\begin{eqnarray}
K({\bf X}_f, {\bf P}_f, t| {\bf X}_0, {\bf P}_0, 0) =  \int \frac{d^N \zeta_f d^N\zeta_0}{(2\pi)^N}  e^{i {\bf P}_{0}\cdot{\bf \zeta}_0 - i {\bf P}_{f} \cdot {\bf \zeta}_f}  \; J({\bf X}_f + \frac{{\bf \zeta}_f}{2}, {\bf X}_f - \frac{{\bf \zeta}_f}{2}, t| {\bf x}_0 + \frac{{\bf \zeta}_0}{2}, {\bf X}_0 - \frac{ {\bf \zeta}_0}{2},0). \label{wfprop}
\end{eqnarray}

Denoting the phase-space coordinates by the vector
\begin{eqnarray}
\xi_a = (X_1, P_1, X_2, P_2, \ldots, X_N, P_N), \hspace{1cm} a = 1, 2, \ldots, 2N, \label{xidef}
 \end{eqnarray}
 we write the Wigner function propagator compactly  as $K_t(\xi_f, \xi_0)$ and express Eq. (\ref{jprop}) as
\begin{eqnarray}
W_t(\xi) = \int \frac{d^{2N} \xi_0}{(2 \pi)^N} K_t(\xi_f, \xi_0) W_0(\xi_0), \label{wt}
\end{eqnarray}
where $W_t$ and $W_0$ are the Wigner functions at times $t$ and $0$, respectively.

\medskip

  In QBM models the Wigner function propagator is Gaussian. This follows from the fact that the total Hamiltonian for the system is quadratic and the initial state for the bath is Gaussian. The most general form of a Gaussian Wigner function propagator is
\begin{eqnarray}
K_t(\xi_f, \xi_0) = \frac{\sqrt{\det S^{-1}(t)}}{\pi^N} \exp \left[ - \frac{1}{2} [\xi_f^a - \xi_{cl}^a(t)] S^{-1}_{ab}(t) [\xi_f^b - \xi_{cl}^b(t)] \right], \label{gauss}
\end{eqnarray}
where $S^{-1}_{ab}(t)$ is a positive real-valued matrix, and $\xi_{cl}(t)$ is the solution of the corresponding classical equations of motion (including dissipation) with  initial condition $\xi = \xi_0$ at $t = t_0$. The equations of motion are linear, so $\xi_{cl}(t)$ is of the form
\begin{eqnarray}
\xi^a_{cl}(t) = R^{a}_b(t) \xi_0^b, \label{ceq}
\end{eqnarray}
in terms of a matrix $R^a_b(t)$.

Equation (\ref{gauss}) holds if there are no ``decoherence-free'' subalgebras, that is, if there is no subalgebra of the canonical variables that remains decoupled from the environment. These observables evolve with a delta-function propagator, rather than with a Gaussian. However, this case corresponds to a set of measure zero in the space of parameters, and it can be obtained as a weak limit
 of the generic expression, Eq. (\ref{gauss}).

\medskip

In order to specify the Wigner function propagator, we must construct the matrix-valued functions  $R(t)$ and $S(t)$. To this end, we consider the two-point  correlation matrix $V$ of a quantum state $\hat{\rho}$, defined by
\begin{eqnarray}
V_{ab} = \frac{1}{2} Tr\left[\hat{\rho} (\hat{\xi}_a \hat{\xi}_b + \hat{\xi}_b \hat{\xi}_a) \right] - Tr (\hat{\rho} \hat{\xi}_a) Tr(\hat{\rho} \hat{\xi}_b). \label{Vab}
\end{eqnarray}

Gaussian propagation decouples the evolution of two-point correlations from any higher-order correlations. From Eqs. (\ref{wt}) and (\ref{gauss}), we find the two-point correlation matrix, Eq. (\ref{Vab}), $V_t$ at time $t$,
\begin{eqnarray}
V_t = R(t)V_0 R^T(t) +  S(t), \label{Vt}
\end{eqnarray}
where $V_0$ is the correlation matrix of the initial state.
 The first term in the right-hand side of Eq. (\ref{Vt}) corresponds to the evolution of the initial phase-space correlations according to the {\em classical} equations of motion. The second term
incorporates the effect of environment-induced fluctuations and it does not
  depend on the initial state. Hence, the matrix  $S$ can be explicitly constructed, by identifying the part of the correlation matrix that does not depend on the initial state.

To this end, we proceed as follows. From the Heisenberg-picture evolution of the bath oscillators, we obtain the equations
\begin{eqnarray}
\ddot{\hat{q}}_i (t) + \omega_i^2 \hat{q}_i (t) =  \sum_r \frac{c_{ir}}{m_i} \hat{X}_r(t), \label{qeq}
\end{eqnarray}
with solution
\begin{eqnarray}
\hat{q}_i(t) = \hat{q}_i^0(t) + \sum_r\frac{c_{ir}}{m_i \omega_i} \int_0^t ds \sin\left(\omega_i(t-s)\right) \hat{X}_r(s), \label{q1}
\end{eqnarray}
where
\begin{eqnarray}
\hat{q}_i^0(t) = \hat{q}_i \cos \left(\omega_i t\right) + \frac{\hat{p}_i}{m_i \omega_i} \sin\left(\omega_it\right).
\end{eqnarray}
For the system variables, we obtain
\begin{eqnarray}
\ddot{\hat{X}}_r (t) + \Omega_r^2 \hat{X}_r (t) + \frac{2}{M_r} \sum_{r'} \int_0^t ds \gamma_{rr'}(t-s) \hat{X}_{r'}(s) = \sum_i \frac{c_{ir}}{M_r} \hat{q}^0_i(t) \label{Xeq},
\end{eqnarray}
where
\begin{eqnarray}
\gamma_{rr'} (s) = - \sum_i \frac{c_{ir} c_{ir'}}{2 m_i \omega_i^2} \sin\left(\omega_i s\right)
\end{eqnarray}
is   the dissipation kernel.
In general, the matrix $\gamma_{rr'}$ is symmetric and has $\frac{1}{2}N(N+1)$ independent terms, each defining a different relaxation time-scale for the system. However, symmetries of the couplings $c_{ir}$ may reduce the number of independent components of the dissipation kernel.

The solution of Eq. (\ref{Xeq}) is
\begin{eqnarray}
\hat{X}_r(t) = \sum_{r'} (\dot{v}_{rr'}(t) \hat{X}_{r'} + \frac{1}{M_{r'}}v_{rr'} \hat{P}_{r'}) +\sum_{r'} \frac{1}{M_{r'}} \int_0^t ds v_{r r'}(t-s) \sum_i c_{ir'} \hat{q}_i^0(s), \label{Xsol}
\end{eqnarray}
where $v_{rr'}(t)$ is the solution of the homogeneous part of Eq. (\ref{Xeq}), with initial conditions $v_{rr'}(0) = \delta_{rr'} $ and $\dot{v}_{rr'}(0) = 0 $. It can be expressed as an inverse Laplace transform
\begin{eqnarray}
v(t) = {\cal L}^{-1} [A^{-1}(z)], \label{vt}
\end{eqnarray}
where $A_{rr'}(z) = (z^2 + \Omega_r^2)\delta_{rr'} + \frac{2}{M_r} \tilde{\gamma}_{rr'}(z)$ and $\tilde{\gamma}_{rr'}(z)$ is the Laplace transform of the dissipation kernel.

The classical equations of motion follow from the expectation values of $\hat{X}_r$ and $\hat{P}_r = M_r \dot{X}_r$ in Eq. (\ref{Xsol}) \begin{eqnarray}
\left(\begin{array}{c} X(t) \\ P(t) \end{array} \right) = \left( \begin{array}{cc} \dot{v}(t) &  v(t) M^{-1}\\  M \dot{v}(t) & M \ddot{v}(t) M^{-1} \end{array} \right) \left(\begin{array}{c} X(0) \\ P(0) \end{array} \right), \label{ceq2}
\end{eqnarray}
where $M = \mbox{diag} ( M_1, \ldots, M_r)$ is the mass matrix for the system. The matrix $R$  of Eq. (\ref{ceq}) follows from Eq. (\ref{ceq2}) by a relabeling coordinated according to the definition of the vector $\xi^a$, Eq. (\ref{xidef}).

We next  employ Eq. (\ref{Xsol}), in order to construct the correlation matrix Eq. (\ref{Vab}). Using the following equation for the correlation functions of harmonic oscillators in a thermal state at temperature $T$,
\begin{eqnarray}
\langle \hat{q}_i^0(s) \hat{q}_j^0(s') \rangle_{T} = \delta_{ij} \frac{1}{2m_i\omega_i} \coth \left(\frac{\omega_i}{2T}\right) \cos \left(\omega_i(s-s')\right),
\end{eqnarray}
we find
\begin{eqnarray}
S_{X_r X_{r'}} &=& \sum_{qq'} \frac{1}{M_q M_{q'}} \int_0^t ds \int_0^t ds' v_{rq}(s) \nu_{qq'}(s-s')  v_{q'r'}(s'),\label{sxx} \\
S_{P_r P_{r'}} &=& M_r M_{r'} \sum_{qq'} \frac{1}{M_q M_{q'}} \int_0^t ds \int_0^t ds' \dot{v}_{rq}(s) \nu_{qq'}(s-s')  \dot{v}_{q'r'}(s'),\\
S_{X_r P_{r'}} &=& M_{r'} \sum_{qq'} \frac{1}{M_q M_{q'}} \int_0^t ds' v_{rq}(s) \nu_{qq'}(s-s')  \dot{v}_{q'r'}(s') \label{sxp},
\end{eqnarray}
where the symmetric matrix
\begin{eqnarray}
\nu_{rr'}(s) = \sum_i \frac{c_{ir} c_{ir'}}{2 m_i \omega_i^2} \coth \left( \frac{\omega_i}{2T}\right)\cos \left(\omega_is\right)
\end{eqnarray}
is the noise kernel. Similarly to the dissipation kernel, the noise kernel has $\frac{1}{2}N(N+1)$ independent components.

\medskip

Equations (\ref{sxx}---\ref{sxp}) together with the classical equations of motion (\ref{ceq2}) fully specify the Wigner function propagator. The master equation in the Wigner representation easily follows,
 by taking the time derivative of Eq. (\ref{wt}) and using the identities
\begin{eqnarray}
 \int \frac{d^{2N} \xi_0}{(2 \pi)^N} (\xi - \xi_{cl})^a K_t(\xi_f, \xi_0) W_0(\xi_0) &=& - S^{ab} \frac{\partial W_t(\xi)}{\partial \xi^b}, \\
  \int \frac{d^{2N} \xi_0}{(2 \pi)^N} (\xi - \xi_{cl})^a (\xi - \xi_{cl})^b K_t(\xi_f, \xi_0) W_0(\xi_0) &=&  S^{ab} + S^{ac}S^{bd} \frac{\partial^2 W_t(\xi)}{\partial \xi^c \partial \xi^d}.
\end{eqnarray}

The result is
\begin{eqnarray}
\frac{\partial W_t}{\partial t} = -(\dot{R}R^{-1})^a_b \frac{\partial (\xi^b W_t)} {\partial \xi^a} + (\frac{1}{2} \dot{S}^{ab} - (\dot{R}R^{-1})_c^{(a} S^{cb)}) \frac{\partial^2 W_t(\xi)}{\partial \xi^a \partial \xi^b}. \label{master}
\end{eqnarray}

The method leading to the master equation (\ref{master}) is a generalization of the approach in Ref. \cite{HalYu} for the derivation of the Hu, Paz and Zhang master equation for $N = 1$. To the best of our knowledge the only other derivation of the QBM master equation in such a general setup (also including external force terms)  is the one by Fleming, Roura and Hu, Ref. \cite{QBM2}. The benefit of the present derivation is that, by construction, it also provides the solution of the master equation, i.e., explicit formulas for the coefficients of the propagator.

 The first term in the right-hand side of Eq. (\ref{master}) corresponds to the Hamiltonian and dissipation terms, and the second one to diffusion with diffusion functions $D^{ab}(t) =  (\frac{1}{2} \dot{S}^{ab} - \dot{R}R^{-1})_c^{(a} S^{cb)})$.
A necessary condition for the master equation to be Markovian is that dissipation is local, that is, that the matrix $A:= \dot{R}R^{-1}$ is time independent. Then $A$ is a generator of a one-parameter semi-group on the classical-state space.  Moreover, the diffusion functions must be constant, which implies that $S$ must be a solution of the equations $\ddot{S} = O\dot{S}+\dot{S}O$.

\section{Generalized uncertainty relations}
In this section, we derive generalized uncertainty relations for the QBM models described in Sec. II, which are relevant to the discussion of entanglement dynamics.

\subsection{Background}

Let ${\cal H} = L^2(R^N)$ be the Hilbert space of a quantum system corresponding to a classical phase-space $R^{2N}$. ${\cal H}$  carries a representation of canonical commutation relations
\begin{eqnarray}
[\hat{q}_i, \hat{p}_j] = i \delta_{ij}, i = 1, \ldots, N.
\end{eqnarray}
 We employ a vector notation, analogous to  Eq. (\ref{xidef}), for the canonical operators $\hat{q}_i$ and $\hat{p}_j$. Then the  commutation relations take the form
\begin{eqnarray}
[\hat{\xi}_a, \hat{\xi}_b] = i \Omega_{ab}, \hspace{0.5cm} a,b = 1,2, \ldots 2N,
\end{eqnarray}
where
\begin{eqnarray}
\Omega = \left(\begin{array}{cccc} J &0 & \ldots  &0 \\
                               0& J & \ldots &0 \\
                               \ldots& & &\\
                               0& 0& \ldots &J \end{array} \right) \hspace{1.2cm}
                               J = \left(\begin{array}{cc} 0&1 \\ -1&0\end{array} \right).
\end{eqnarray}

The standard uncertainty relations for this system take the form
\begin{eqnarray}
V \geq - \frac{i}{2} \Omega. \label{V}
\end{eqnarray}

 For  a bipartite system,   with $n$ degrees of freedom for the first subsystem, and   $N-n$ ones for the second, the  Peres-Horodecki partial transpose operation defines a transformation $\xi \rightarrow \Lambda \xi$, where $\Lambda$ inverts the momenta of the second subsystem. Then, the correlation matrix of a separable state  satisfies the inequality \cite{Simon}
\begin{eqnarray}
V \geq - \frac{i}{2} \tilde{\Omega}, \hspace{1cm} \tilde{\Omega} = \Lambda \Omega \Lambda. \label{V2}
\end{eqnarray}

\medskip
Of special interest is the case $N = 2$, where Eqs. (\ref{V}) and (\ref{V2}) lead to a simple, if weaker, set of uncertainty relations. These have a simple generalization in the QBM model considered in this article. We introduce  the variables
\begin{eqnarray}
X_+ = \frac{1}{2}(X_1 + X_2) , \hspace{1cm} P_+ = P_1 + P_2 , \\
X_- = \frac{1}{2}(X_1 - X_2) , \hspace{1cm} P_- = P_1 - P_2.
\end{eqnarray}
The partial transpose operation then interchanges $P_+$ with $P_-$, that is,
\begin{eqnarray}
\Lambda (X_+, P_+, X_-, P_-) = (X_+, P_-, X_- , P_+).
\end{eqnarray}

Hence, the uncertainty relations,
 \begin{eqnarray}
{\cal A}_{X_+P_+} :=  (\Delta X_+)^2 (\Delta P_+ )^2 - V_{X_+P_+}^2 \geq \frac{1}{4},  \hspace{0.8cm}
{\cal A}_{X_-P_-} := (\Delta X_-)^2 (\Delta P_- )^2 - V_{X_-P_-}^2 \geq \frac{1}{4} \label{unc2a},
\end{eqnarray}
satisfied by any pair of conjugate variables (they follow from the positivity of  the $2\times 2$ diagonal subdeterminants of $V$), imply that a factorized state must satisfy the following relations
\begin{eqnarray}
{\cal A}_{X_+ P_-} := (\Delta X_+)^2 (\Delta P_- )^2 - V_{X_+ P_-}^2 \geq \frac{1}{4}, \hspace{0.8cm}
{\cal A}_{X_- P_+} :=  (\Delta X_-)^2 (\Delta P_+ )^2 - V_{X_-P_+}^2 \geq \frac{1}{4} \label{unc2b}.
 \end{eqnarray}
If either inequality in Eq. (\ref{unc2b}) is violated, then the state is entangled. Hence, the uncertainty functions ${\cal A}_{X_+ P_-}$ and ${\cal A}_{X_- P_+}$ provide witnesses of entanglement for any state. They are weaker than the full Eq. (\ref{V2}). Equation (\ref{V2}) fully specifies entanglement in all Gaussian states, while Eq. (\ref{unc2b}) does so only for pure Gaussian states.

\subsection{Uncertainty relations in QBM models}

 The initial correlation matrix $V_0$  in  Eq. (\ref{Vt}) satisfies
  the inequality (\ref{V}). It follows that
 \begin{eqnarray}
 V_t \geq - \frac{i}{2} R(t) \Omega R^T(t) + S(t). \label{gunc1}
 \end{eqnarray}
The inequality (\ref{gunc1}) is a generalized uncertainty relation that incorporates the effect of environment-induced fluctuations. It generalizes the uncertainty relations of Ref. \cite{AnHa} to oscillator systems with an arbitrary number of degrees of freedom. The right-hand side of Eq. (\ref{gunc1}) depends only on the coefficients of the Wigner function propagator and not on any properties of the initial state. Hence, Eq. (\ref{gunc1}) provides a lower bound to the correlation matrix  at time $t$, for a system that comes into contact with a heat bath at time $t = 0$.

 Equality in Eq. (\ref{gunc1}) is achieved for pure Gaussian states. The bound is to be understood in the sense of an envelope. No single Gaussian state saturates the bound in Eq. (\ref{gunc1}) at all moments of time, but equality is achieved by a different family of Gaussians at each moment $t$.

\subsubsection{Bipartite entanglement}
When applied to a bipartite system, Eq. (\ref{gunc1}) implies that the condition
\begin{eqnarray}
- \frac{i}{2} R(t) \Omega R^T(t) + S(t) < -\frac{i}{2} \tilde{\Omega} \label{gunc2}
\end{eqnarray}
 is sufficient for the existence of entangled states at time $t$, irrespective of the degradation caused by the environment. For Gaussian initial states,  this condition is also necessary.


For a factorized initial state, Eqs. (\ref{Vt}) and (\ref{V2}) yield
\begin{eqnarray}
 V_t \geq - \frac{i}{2} R(t) \tilde{\Omega} R^T(t) + S(t). \label{gunc3}
\end{eqnarray}

Inequality (\ref{gunc3}) is saturated for {\em factorized} pure Gaussian states, and, similarly to Eq. (\ref{gunc3}), the lower bound to the correlation matrix is to be understood as an envelope.

 If an initially factorized state remains factorized at time $t$, then $V_t \geq -\frac{i}{2}\tilde{\Omega}$. Then Eq. (\ref{gunc3}) implies that the inequality
 \begin{eqnarray}
 - \frac{i}{2} \left( R(t) \tilde{\Omega} R^T(t) - \tilde{\Omega} \right) + S(t) \leq 0 \label{gunc4}
 \end{eqnarray}
 is a necessary condition for the preservation of factorizability at time $t$.

\subsubsection{Tripartite entanglement}
Equations (\ref{Vt}) and (\ref{gunc1}) apply to systems of $N$ oscillators. Used in conjunction with suitable separability criteria for multipartite systems \cite{DCT}, they also allow the derivation for uncertainty relations relevant multipartite systems. For example, we can use the criteria of Ref. \cite{GKLC} which apply to systems of three oscillators, labeled by the index $i = 1, 2, 3$. One defines the matrices $\Lambda_i$ that effect partial transposition with respect to the $i$th subsystems. Then, separable states satisfy 
\begin{eqnarray}
V \geq - \frac{i}{2} \tilde{\Omega}_i, \hspace{1cm} \tilde{\Omega}_i = \Lambda \Omega_i \Lambda, \label{unc6}
\end{eqnarray}
for all $i$. There are some subtleties in the application of the criterion Eq. (\ref{unc6}) for Gaussians: there exist states that satisfy Eq. (\ref{unc6}) that are not fully separable, but only biseparable with respect to all possible bipartite splits---see Ref. \cite{GKLC} for details. However, the reasoning of Sec. III B 1 applies. 

The condition 
\begin{eqnarray}
- \frac{i}{2} R(t) \Omega R^T(t) + S(t) < -\frac{i}{2} \tilde{\Omega}_i, \label{guncb1}
\end{eqnarray}
for all $i$, is sufficient for the existence of entangled states at time $t$, irrespective of the degradation caused by the environment. For a factorized initial state, Eqs. (\ref{Vt}) and (\ref{unc6}) yield
\begin{eqnarray}
 V_t \geq - \frac{i}{2} R(t) \tilde{\Omega}_i R^T(t) + S(t), \label{guncb2}
\end{eqnarray}
for all $i$. Equation (\ref{guncb2}) implies that the condition 
 \begin{eqnarray}
 - \frac{i}{2} \left( R(t) \tilde{\Omega} R^T(t) - \tilde{\Omega}_i \right) + S(t) \leq 0 \label{guncb3}
 \end{eqnarray}
 is necessary for the preservation of factorizability at time $t$.

\subsection{A case model}

The uncertainty relations (\ref{gunc1})--(\ref{gunc4}) hold for any Gaussian QBM system and depend only on the matrices $R$ and $S$ defining  the density-matrix propagator, for which explicit expressions were given in Sec. II. In what follows, we elaborate on these relations in the context of  a specific QBM model for a bipartite system, which has been studied by Paz and Roncanglia \cite{PR1, PR2}.

 In this model, the system consists of  two harmonic oscillators with equal masses $M$ and frequencies $\Omega_1$ and $\Omega_2$. We also consider symmetric coupling to the environment, that is., $c_{i1} = c_{i2}:= c_i$ in Eq. (\ref{hint}). The latter assumption is a strong simplification, because
the dissipation and noise kernels then  become scalars,
\begin{eqnarray}
\gamma(s) &=& \int d \omega I(\omega) \sin \left(\omega s\right) \left(\begin{array}{cc}1&1\\1&1 \end{array} \right), \\
\nu(s) &=& \int d \omega I(\omega) \coth \left(\frac{\omega}{2T}\right) \cos \left(\omega s\right) \left(\begin{array}{cc}1&1\\1&1 \end{array} \right),
\end{eqnarray}
where
\begin{eqnarray}
I(\omega) = \sum_i \frac{c_i^2}{2 m_i \omega_i^2} \delta (\omega - \omega_i)
\end{eqnarray}
is the bath's spectral density. A common form for $I(\omega)$  is

\begin{eqnarray}
I(\omega)=M\gamma\omega \left(\frac{\omega}{\tilde{\omega}} \right)^s e^{-\frac{\omega^2}{\Lambda^2}},
\end{eqnarray}
where $\gamma$ is a dissipation constant, $\Lambda$ is a high-frequency cutoff, $\tilde{\omega}$ is a frequency scale, and the exponent $s$ characterizes the infrared behavior of the bath. For this model, it is convenient to employ the dimensionless parameter $\delta := \frac{\Omega_1^2 - \Omega_2^2}{ \Omega_1^2 + \Omega_2^2 }$, denoting how far the system is from resonance, and the scaled temperature $\theta := \frac{T}{\sqrt{ \Omega_1^2 + \Omega_2^2 }}$.

 In this model, the pair of oscillators is coupled to the environment only through the variables $X_+$. The variable $X_-$ is affected by the environment only through its coupling with $X_+$, which is proportional to $\Delta^2 = |\Omega_i^2 - \Omega_2^2|$.  For resonant oscillators ($\Omega_1 = \Omega_2$) this coupling vanishes, the subalgebra generated by $\hat{X}_-$ and $\hat{P}_-$ is isolated from the environment, and it is therefore decoherence free. This means in particular that some entanglement may persist even at late times. This case has been studied in detail in Ref. \cite{PR1}. For  nonzero $\Delta$,  the $\hat{X}_-$ and $\hat{P}_-$ subalgebra is not totally isolated from the environment.

\medskip

 The uncertainty relations simplify when the environment is ohmic ($s = 0$). Then, dissipation is local and in the weak-coupling limit ($\gamma << \Omega_i$), the matrices $R$ describing classical evolution take the form

  \begin{eqnarray}
  R(t) = e^{- \frac{1}{2}\gamma t} U(t),
  \end{eqnarray}
 where $U(t)$ is a canonical transformation: $U(t)\Omega U^T(t) = \Omega$. Hence, Eq. (\ref{gunc1}) becomes
 \begin{eqnarray}
 V_t \geq -\frac{i}{2} e^{- \gamma t} \Omega + S(t). \label{gunc5}
 \end{eqnarray}
From Eq. (\ref{gunc5}) we see that dissipation tends to shrink phase-space areas, but this is compensated by the effects of diffusion incorporated into the definition of the matrix $S$. For an initial factorized state, we obtain
\begin{eqnarray}
V_t \geq -\frac{i}{2} e^{- \gamma t} F(t) + S(t), \label{gunc6b}
\end{eqnarray}
where $F(t) = U(t)\tilde{\Omega}U^T(t)$ is an oscillating function of time. The oscillations in $F(t)$ may lead to violation of the bound $V_t \geq -\frac{i}{2} \tilde{\Omega}$ for factorized states and thus to entanglement creation. However, the oscillating character of $F(t)$ implies that  entanglement creation will in general be accompanied by entanglement death and revival. For times  $t >> \gamma^{-1}$ the first term in the right-hand side of Eq. (\ref{gunc6b}) is suppressed.

\paragraph{The Wigner function area.} According to Eq. (\ref{gunc5}), the matrix $V_t + \frac{i}{2}e^{- \gamma t} -  S(t)$ is positive. Its upper $2\times 2$ submatrix in the $X_+, X_-$ coordinates should also be positive; hence,
  \begin{eqnarray}
  [(\Delta X_+)^2 -  S_{X_+X_+}][(\Delta P_+)^2 -  S_{P_+P_+}]] - (V_{X_+P_+} -  S_{X_+P_+})^2 \geq \frac{1}{4} e^{- 2 \gamma t}. \label{gunc6}
  \end{eqnarray}
By virtue of Schwartz's inequality, $(\Delta X_+)^2 S_{X_+X_+} + (\Delta P_+)^2 S_{P_+P_+} - V_{X_+P_+}S_{X_+P_+} \geq 0$; hence,
\begin{eqnarray}
{\cal A}_{X_+P_+} \geq \frac{1}{4} e^{-  \gamma t} +  \left(S_{X_+X_+} S_{P_+P_+} - S_{X_+P_+}^2\right ). \label{gunc7a}
\end{eqnarray}
Similarly,
\begin{eqnarray}
{\cal A}_{X_- P_-} &\geq& \frac{1}{4} e^{- \gamma t} + \left(S_{X_-X_-} S_{P_- P_-} - S_{X_- P_-}^2\right), \label{gunc7b} \\
{\cal A}_{X_+ P_-} &\geq&  \left(S_{X_+X_+} S_{P_- P_-} - S_{X_+ P_-}^2 \right),
\label{gunc7c}\\
{\cal A}_{X_-, P_+} &\geq&  \left(S_{X_-X_-} S_{P_+P_+} - S_{X_- P_+}^2 \right).
\label{gunc7d}
\end{eqnarray}

The uncertainty functions ${\cal A}_{X_iP_j}$ correspond to the area of the projection of the Wigner function ellipse onto a two-dimensional subspace defined by $X_i$ and $P_j$. The right-hand side of the inequalities are plotted in Fig. 1 as function of time. Except possibly at early times, the functions increase monotonically and reach a constant asymptotic value at a time scale of order $\gamma^{-1}$.

   \begin{figure}[tbp]
   \centering
\includegraphics[width=0.9 \textwidth]{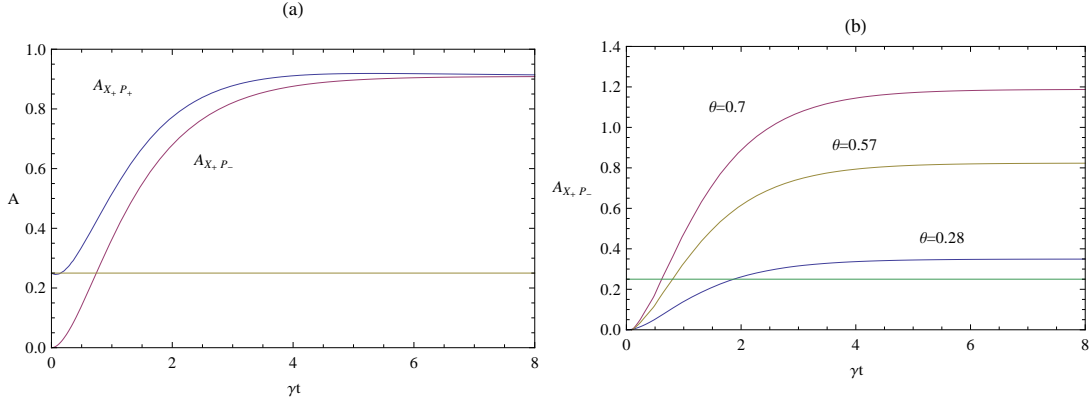} \caption{ \small (Color online) (a) The lower bounds for ${\cal A}_{X_+P_-}$ in Eq. (\ref{gunc7a}) and ${\cal A}_{X_+P_+}$ in Eq. (\ref{gunc7c}) as  functions of $\gamma t$, for parameter values $\theta = 0.7$ and $\delta = 0.38$. (b) The lower bound to ${\cal A}_{X_+P_-}$ as a function of $\gamma t$ for different values of the dimensionless temperature $\theta$.}
\end{figure}




\section{Entanglement dynamics}

\subsection{Disentanglement at high temperature}

 A widely studied regime in quantum Brownian motion models is the so-called Fokker-Planck limit in ohmic environments, because in this limit the master equation is Markovian. The Fokker-Planck limit is defined by the condition  $T >> \Lambda$, and then taking $\Lambda \rightarrow \infty$, in order to obtain time-local dissipation and noise.

In this regime, thermal noise is strong, resulting in loss of  quantum coherence and entanglement at early times. It is convenient to work with the uncertainty functions  ${\cal A}_{X_iP_j}$, because they can be explicitly evaluated\footnote{There is no loss of information in this choice, because of the rapid degradation of coherence. The sharper inequality, Eq. (\ref{gunc5}), gives the same estimation for the characteristic time scales of these processes.}.
We find
\begin{eqnarray}
{\cal A}_{X_+P_+} &\geq& \frac{1}{4}(1 - \gamma t + \gamma^2 T^2 t^4), \label{gunc9a}
 \\
{\cal A}_{X_-P_-} &\geq& \frac{1}{4}[1 - \gamma t + \frac{\gamma^2 T^2}{2^{8}\cdot 3^4\cdot 35} \Delta^8 t^{12}], \label{gunc9b}
\\
{\cal A}_{X_+P_-} &\geq&  \frac{11 \gamma^2 T^2}{256} \Delta^4 t^8 , \label{gunc9c}
\\
 {\cal A}_{X_-P_+} &\geq&  \frac{\gamma^2 T^2}{256} \Delta^4 t^8, \label{gunc9d}
\end{eqnarray}
 The above equations are obtained far from resonance for the two oscillators, that is, $\Delta >> \gamma$.

Equations (\ref{gunc9a}) and (\ref{gunc9b}) represent the initial growth of fluctuations starting from purely quantum fluctuations at $t = 0$. The growth of the fluctuations for the variables $X_+$ and $P_+$ is faster than that of the variables $X_-$ and $P_-$, because the former couple indirectly to the bath. The $-\gamma t$ term in these equations indicates an initial decrease of the fluctuations, in apparent violation of the uncertainty principle. The violation in Eq. (\ref{gunc9a}) occurs at  a timescale of order $(\gamma T^2)$. This is because these equations are derived taking the infinite cut-off limit $\Lambda \rightarrow \infty$, which leads to violations of the positivity of the density operator at $t < \Lambda^{-1}$ \cite{AnHa}. For $t > \Lambda^{-1} >> T^{-1}$, and $T$ sufficiently large  so that $\gamma T^2/\Lambda^3 >> 1$,  such violations do not arise.

 Ignoring the positivity-violating terms, Eq. (\ref{gunc9a}) leads to an expression $t_{th} \sim 1/\sqrt{\gamma T}$ for the  time scale where the thermal fluctuations overcome the purely quantum ones. This is an upper limit to the decoherence time for the $X_+$ and $P_+$ variables \cite{AnHa}.

From Eqs. (\ref{gunc9c}) and (\ref{gunc9d}) we obtain the characteristic time scale where ${\cal A}_{X_+ P_-}$ and ${\cal A}_{X_- P_+}$ reach the value $\frac{1}{4}$ starting from 0. This is indicative of the time scale for disentanglement $t_{dis}$ in this model:
\begin{eqnarray}
t_{dis} \sim \frac{1}{(\gamma T \Delta^2)^{1/4}}.
\end{eqnarray}
The characteristic scale for disentanglement is distinct from the time scale $t_{th}$ characterizing the growth of thermal fluctuations:
\begin{eqnarray}
t_{dis}/t_{th} = \left(\frac{\gamma T}{\Delta^2}\right)^{1/4}.
 \end{eqnarray}

 For sufficiently small values of $\Delta$, that is, weak coupling between the $+$ and $-$ variables, the disentanglement timescale may be much larger than the decoherence time scale for the $X_+$ and $P_+$ variables. Hence, even if the $X_-, P_-$ degrees of freedom are only partially protected from degradation  from the environment, they can sustain entanglement
  long after the $X_+$ and $P_+$ variables have decohered.

\subsection{Long-time limit}

While entanglement may be preserved much longer than the coherence of the $X_+$ and $P_+$ degrees of freedom,  the interaction with the environment sets the relaxation time scale $\gamma^{-1}$  as an upper limit for disentanglement time.
For times $t >> \gamma^{-1}$, all states tend toward the stationary state $\hat{\rho}_{\infty}$ corresponding to a Wigner function,
\begin{eqnarray}
W_{\infty}(\xi) = \frac{\sqrt{\det S^{-1}_{\infty}}}{\pi} \exp[- \frac{1}{2} \xi S^{-1}_{\infty} \xi], \label{asympt}
\end{eqnarray}
where $S_{\infty}$ is the asymptotic value of the matrix $S$ at $t \rightarrow \infty$. At this limit, the correlation matrix $V$ coincides with $S$. Explicit evaluation of Eqs. (\ref{sxx}-\ref{sxp})
shows that, as $t\rightarrow \infty$ the only nonvanishing elements of the matrix $S$ are the diagonal ones: $S_{X_+X_+}, S_{X_-X_-}, S_{P_+P_+}, S_{P_-P_-}$ (see the Appendix). For states of this form, the uncertainty functions ${\cal A}_{X_+P_-}$ and ${\cal A}_{X_- P_+}$ fully determine entanglement. We further find that to leading order in $\gamma/\Omega_i$ and $\Omega_i/\Lambda$, the asymptotic state coincides with the thermal state for
 Hamiltonian $\hat{H}_0$ in Eq. (\ref{ho}); hence, it is factorized.

 However, at low temperatures the thermal states are close to the boundary that separates factorized from entangled states (for example, they satisfy ${\cal A}_{X_+P_-} \simeq \frac{1}{4}$). Hence, the corrections from the nonzero values of $\gamma/\Omega_i$ and $\Omega_i/\Lambda$ may lead the asymptotic state to retain some degree of entanglement, as was found in Ref. \cite{PR2}. We have verified numerically that the residual entanglement decreases with increasing values of the cutoff parameter $\Lambda$.

This result applies to a system of  nondegenerate oscillations. For degenerate oscillators, the $X_-$ and $ P_-$ subalgebra is protected from the environment. Hence,  the asymptotic state is not unique and it may sustain entanglement or even be characterized by a nonterminating sequence of entanglement deaths and revivals.

\medskip

The analysis of Sec. II allows us to make a general characterization of  the asymptotic state valid for any QBM model. The key observation is that the  uniqueness of the asymptotic state is solely determined from the classical equations of motion, that is, from the matrix $R^a_{b}$ in Eq. (\ref{ceq}). In the generic case the phase space contains no dissipation-free subspace, and $\xi^a_{cl}(t) \rightarrow 0$ as $t \rightarrow \infty$, irrespective of the initial condition. Hence, for times $t$ much larger than the relaxation time $\tau_{rel}$ the memory of the initial state is lost from the Wigner function propagator, Eq.  (\ref{gauss}). Moreover, if $\xi^a_{cl}(t) \rightarrow 0$ sufficiently fast as $t \rightarrow \infty$, the limit $t \rightarrow \infty$ for the matrix  $S$, Eqs. (\ref{sxx}--\ref{sxp}), is well defined. Thus a unique asymptotic state of the form (\ref{asympt}) is obtained. At the weak-coupling limit, one expects that the asymptotic state will be close to the thermal state at  temperature $T$; hence, it will be factorized.

If, on the other hand, the classical equations of motion admit a dissipation-free subspace, time evolution in this subspace is Hamiltonian, and there $\xi^a_{cl}(t)$ does not converge to a unique value as $t \rightarrow \infty$. This implies that the Wigner function propagator Eq. (\ref{gauss}) preserves its dependence on the initial variables even for $t >> \tau_{rel}$. As a consequence, an asymptotic state may not exist or, if it exists, it may not be unique.  Hence, in this case asymptotic entanglement or a sequence of entanglement death and revivals is possible.

Nonetheless, the case of a unique asymptotic state is the generic one. Dissipation-free subspaces exist only for a set of measure zero in the space of parameters (e.g., system-environment couplings) characterizing a QBM model. For example, even a small dependence of the coupling on the oscillator's position will prevent the existence of a dissipation-free subspace. Hence, unless some symmetry can be invoked that fully protects a subalgebra from degradation from the environment, we expect that the relaxation time sets an absolute upper limit to the time scale that entanglement can be preserved in any oscillator system interacting with a QBM-type environment.

\subsection{Entanglement creation}

In general, two noninteracting quantum systems may become entangled by their interaction with a third system. In QBM the role of the third system can be played by the environment, and indeed, low-temperature baths have the tendency to create entanglement.

 The uncertainty relations Eqs. (\ref{gunc3}) and (\ref{gunc4}) are particularly useful  for the study of entanglement creation. We apply them as follows.
 The positivity of the matrix $V_t + \frac{i}{2} \tilde{\Omega}$ is a necessary criterion for a state to be factorized at time $t$. Hence, in a factorized state, the minimal eigenvalue $\lambda_{min}(t)$ of $V_t + \frac{i}{2} \tilde{\Omega}$ is positive. By Eq. (\ref{gunc3}), $\lambda_{min}(t)$ is always bounded from below by the minimal eigenvalue of the matrix $-\frac{i}{2}(R\tilde{\Omega}R^T - \tilde{\Omega}) + S$, which we denote as $\tilde{\lambda}_{bound}(t)$. Hence, the  function $\tilde{\lambda}_{bound}(t)$ determines the capacity of the environment to create entanglement irrespective of the initial state. In particular, the condition that  $\tilde{\lambda}_{bound}(t) \leq 0$ implies that at least some factorized states can develop entanglement at time $t$.

 Figure 2(b) provides a plot of the minimal eigenvalue $\lambda_{min}(t)$ of $V_t + \frac{i}{2} \tilde{\Omega}$ for an initial factorized Gaussian state together with the lower bound, $\tilde{\lambda}_{bound}(t)$, as functions of time. $\lambda_{min}(t)$  oscillates rapidly at a scale of $\Omega_i^{-1}$, so that at time scale of order $\gamma^{-1}$ we can distinguish only two enveloping curves that bound it from above and below.
  $\tilde{\lambda}_{bound}(t)$ is close to the lower enveloping curve of $\lambda_{min}(t)$ and we note that at specific instants the inequality $\lambda_{min}(t) \geq \lambda_{bound}(t)$ is saturated.

 \begin{figure}[tbp]
   \centering
\includegraphics[width=0.9 \textwidth]{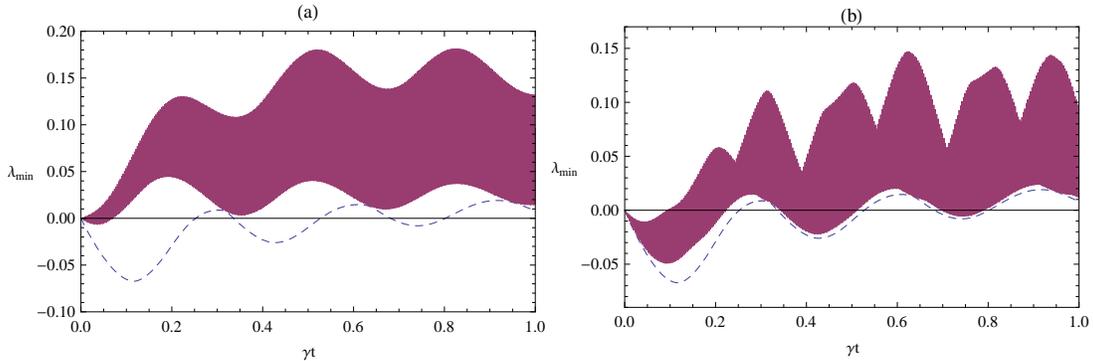} \caption{ \small (Color online) (a) The rapidly oscillating minimal eigenvalue $\lambda_{min}$ of $V_t + \frac{i}{2} \tilde{\Omega}$ for an initial factorized  state  $|0, 1 \rangle$, together with the lower bound $\tilde{\lambda}_{bound}(t)$ corresponding to minimal eigenvalue of the matrix $-\frac{i}{2}(R\tilde{\Omega}R^T - \tilde{\Omega}) + S$. In this plot $\delta = 0.02$ and $\theta = 0.21$.   (b) Same as in (a) but for an initial factorized Gaussian state.}
\end{figure}

  For Gaussian states the criterion $V_t < - \frac{i}{2} \tilde{\Omega}$ completely specifies entanglement, hence, for times $t$ that $\tilde{\lambda}_{bound}(t) > 0$, no initially factorized Gaussian state can sustain entanglement. In Figs. 2(b) and 3, we see that $\tilde{\lambda}_{bound}(t)$ exhibits oscillations around zero at low temperatures. This implies that, at least for Gaussian states, entanglement creation at low temperature is typically accompanied by a period of ``entanglement oscillations'', that is, a sequence of entanglement deaths and revivals, which terminates at a  time scale of order $\gamma^{-1}$, when the system relaxes to an asymptotic factorized state.

 Figure 2(a) provides a plot  of $\lambda_{min}(t)$  for an initial factorized energy eigenstate $|0, 1\rangle$, together with the bound $\tilde{\lambda}_{bound}(t)$. For non-Gaussians, a positive value of $\lambda_{min}(t)$ does not imply factorizability of the state; information about entanglement is carried in higher order correlation functions of the system. Nonetheless, a negative value of $\lambda_{min}$
  is a definite sign of entanglement. Despite of the fact that $\lambda_{min}(t)$ saturates the bound at some instants, in general its behavior is qualitatively different.

 In Fig. 3, the minimal eigenvalue $\tilde{\lambda}_{bound}(t)$ is plotted for different values of temperature. With increasing temperature the time intervals of persisting entanglement shrink and the entanglement oscillations are suppressed. At sufficiently high temperature (of order $\theta > 10$), no creation of entanglement occurs.

 \begin{figure}[tbp]
   \centering
\includegraphics[width=0.5 \textwidth]{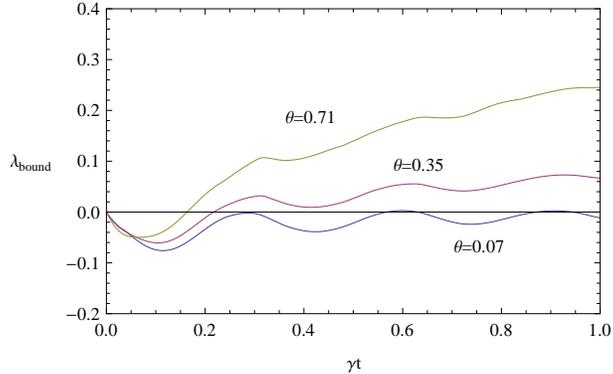} \caption{\small (Color online) The minimal eigenvalue $\tilde{\lambda}_{bound}(t)$ of the  matrix $-\frac{i}{2}(R\tilde{\Omega}R^T - \tilde{\Omega}) + S$ for $\delta = 0.02$ and different values of temperature.}
\end{figure}

\subsection{Disentanglement at low temperature}
We saw that at high temperature, the noise from the environment  degrades the quantum state and causes rapid decoherence and disentanglement. At low temperatures ($\theta < 1 $), however, the noise is not sufficiently strong to cause decoherence \cite{HPZ}, and entanglement is preserved longer. The physical mechanism responsible for disentanglement at low-temperature is relaxation: the existence of a unique asymptotic factorized state implies that at a time scale of order $\gamma^{-1}$ all memory of the initial state (including entanglement) is lost. In other words, a low temperature bath is much more efficient in creating and preserving entanglement, but relaxation to equilibrium will inevitably lead to a factorized state.

  By Eq. (\ref{gunc1}), the minimal  eigenvalue $\lambda_{min}(t)$ of the matrix $V_t + \frac{i}{2} \tilde{\Omega}$ is always bounded from below by the minimal eigenvalue $\lambda_{bound}(t)$ of the matrix $-\frac{i}{2}(R\Omega R^T - \tilde{\Omega}) + S$. Hence, the condition $\lambda_{bound}(t) < 0$ is sufficient for the existence of entangled states at time $t$. Moreover, the condition $\lambda_{min}(t) > 0 $ establishes that the evolution of any {\em Gaussian} initial state at time $t$ is factorized.

Figure 4 contains  plots of the minimal eigenvalue $\lambda_{min}(t)$ of $V_t + \frac{i}{2} \tilde{\Omega}$ for two different initial states, together with the lower bound  $\lambda_{bound}(t)$. In Fig. 4(a) the initial state is an entangled Gaussian, and in Fig. 4(b) the initial state is $\frac{1}{2(1 + e^{- |z|^2})} (|z, 0\rangle +|0, z\rangle)$, where $z$ is a coherent state. In both cases, $\lambda_{min}(t)$ approaches the lower bound only after a time scale of order $\gamma^{-1}$ when the system has started relaxation to a unique asymptotic state. We note that there are no entanglement oscillations for such states, only a gradual decay of entanglement. This behavior is typical for initial states that violate Eq. (\ref{V2}) by a substantial margin. However, the uncertainty relations do not provide any significant information about the entanglement dynamics of  initial states that are entangled, but do not violate the bound, Eq. (\ref{V2}). This is the case, for example, for states of the form $\frac{1}{\sqrt{2}}(|0,1\rangle + e^{i \theta} |1, 0\rangle))$. In order to  study such states, we would have to obtain generalized uncertainty relations pertaining to correlation functions of order higher than 2.

 \begin{figure}[tbp]
   \centering
\includegraphics[width=0.9  \textwidth]{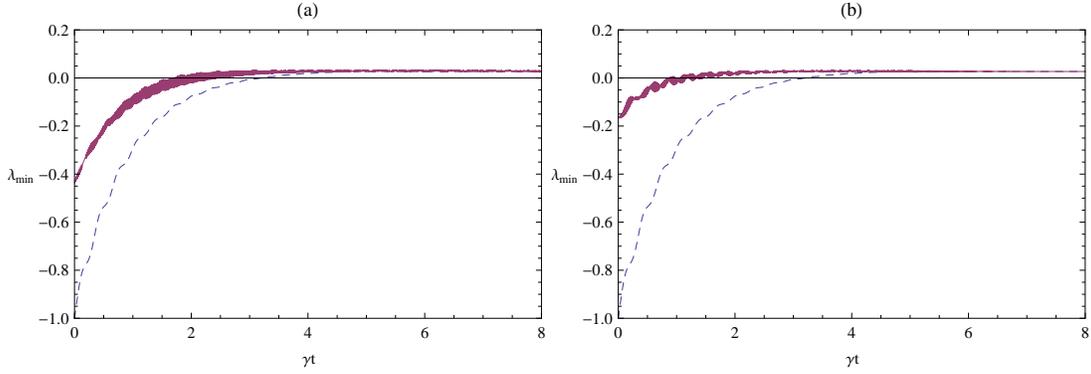} \caption{ \small (Color online) (a) The rapidly oscillating minimal eigenvalue of $V_t + \frac{i}{2} \tilde{\Omega}$ for an initial entangled Gaussian state, together with the lower bound corresponding to minimal eigenvalue of the matrix $-\frac{i}{2}(R\Omega R^T - \tilde{\Omega}) + S$. In this plot $\delta = 0.02$ and $\theta = 0.21$.   (b) Same as in (a) but for the initial state $\frac{1}{2(1 + e^{- |z|^2})} (|z, 0\rangle +|0, z\rangle)$. }
\end{figure}

In Fig. 5, we plot the minimal eigenvalue $\lambda_{bound}(t)$ as a function of time $t$, for different temperatures. As expected, the time interval during which the system sustains entangled states [i.e., $\lambda_{bound}(t) < 0$ ] shrinks with temperature.

 \begin{figure}[tbp]
   \centering
\includegraphics[width=0.5 \textwidth]{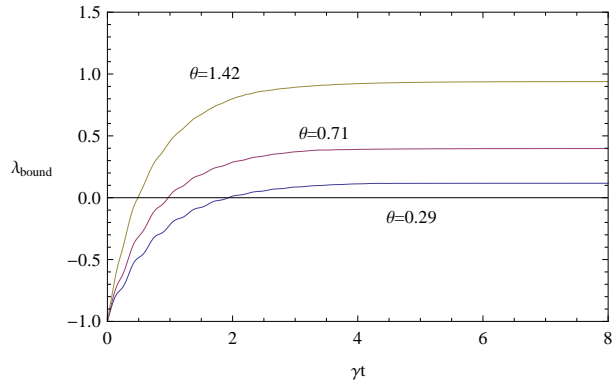} \caption{ \small (Color online) The minimal eigenvalue $\lambda_{bound}(t)$ of the  matrix $-\frac{i}{2}(R\Omega R^T - \tilde{\Omega}) + S$ for $\delta = 0.02$ and different values of temperature.}
\end{figure}

The uncertainty relation, Eq. (\ref{gunc1}), allows for the definition of
the disentanglement time $t_{dis}$ as the instant that $\lambda_{bound}(t) = 0$. Thus defined, $t_{dis}$ is an upper bound to the disentanglement time for {\em any} Gaussian initial state. In general, non-Gaussian states may preserve entanglement for times larger than $t_{dis}$. However, $t_{dis}$ depends only on the matrices $S$ and $R$, and
the evolution of higher-order correlation functions of non-Gaussian states is governed by the matrices $S$ and $R$ alone. Moreover, $t_{dis} \sim \gamma^{-1}$ refers to the regime of relaxation to a unique thermal equilibrium state, hence, the loss of any memory of the initial condition.
 For this reason,  it is reasonable to assume that  $t_{dis}$ provides a good estimation for disentanglement time that is valid for a larger class of initial states, at least as far as its qualitative dependence on temperature and other bath parameters are concerned. Figure 6 plots $t_{dis}$ as a function of temperature for different values of $\delta$. As expected $t_{dis}$ decreases with temperature. However, there is no monotonic dependence of $t_{dis}$ on $\delta$, and for $\theta > 0.5$, $t_{dis}$ is largely insensitive to $\delta$.

Finally, we note that the  weaker uncertainty relations for the Wigner function areas ${\cal A}_{X_{\pm}P_{\mp}}$ also provide an estimation for disentanglement time $t_{is}$. Since these inequalities are weaker, the values of $t_{dis}$ thus obtained are smaller,   but their dependence on the parameters $\delta$ and $\theta$ is qualitatively similar.

\begin{figure}[tbp]
   \centering
\includegraphics[width=0.5 \textwidth]{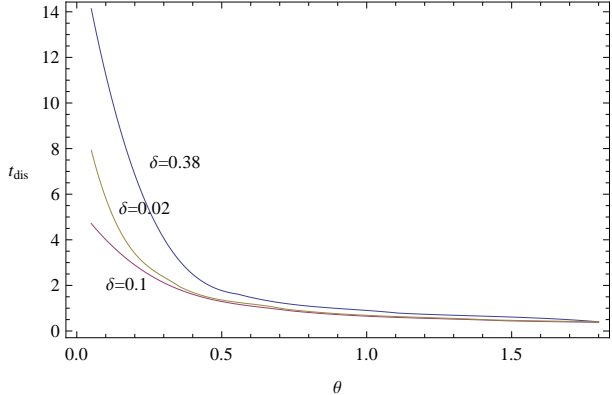} \caption{\small (Color online) Disentanglement time $t_{dis}$ in units of $\gamma^{-1}$ as a function of the dimensionless temperature $\theta$ for different values of $\delta$.}
\end{figure}

\section{Conclusions}
The main results of our article are the following: (i) the explicit construction of the Wigner function propagator for  QBM models with any number of system oscillators and for any spectral density; the propagator allows for a simple derivation of the corresponding master equation; (ii) the identification of generalized uncertainty relations valid in any QBM model that provide a state-independent lower bound to the fluctuations induced by the environment; (iii) the application of the uncertainty relations to a concrete model, for the study of decoherence, disentanglement, and entanglement creation in different regimes. In particular, we showed that entanglement creation is often accompanied by entanglement oscillation at early times and that the uncertainty relations provide an upper bound to disentanglement time with respect to all initial Gaussian states.

In our opinion, the most important feature of the techniques developed in this article is that they can be immediately generalized for addressing more complex systems and issues in the study of entanglement dynamics, for example, in the derivation of uncertainty relations for higher-order correlation functions, or for information-theoretic quantities that contain more detailed information about entanglement of general initial states, and in the exploration of entanglement dynamics in multipartite systems and of the dependence of entanglement on the spatial separation of multipartite systems.

\begin{appendix}
\section{The coefficients in the Wigner function propagator}
In this appendix, we sketch the calculations of the coefficients in the Wigner function propagator for the model presented in Sec. III C.

We first compute the function $v(s)$ of Eq. (\ref{vt}) in the $X_+, X_-$ coordinates. To leading order in $\gamma$ for the poles in the Laplace transform (\ref{vt}), we obtain

\begin{eqnarray}
v_{++}(s)&=&\frac{e^{-\frac{1}{2}\gamma s}}{4\Omega_1 \Omega_2(\Omega_2^2-\Omega_1^2)}
\left[\Omega_2 \sin(\Omega_1 s)\left(\gamma^2 - 2\Omega_1^2 + 2 \Omega_2^2 \right) \right. 
\nonumber \\
&&\left. + 4 \gamma \Omega_1\Omega_2 \left(\cos(\Omega_2 s)-\cos(\Omega_1 s)\right)-\Omega_1 \sin(\Omega_2 s) \left(\gamma^2+2\Omega_1^2-2 \Omega_2^2\right)\right]
\\
v_{+-}(s)&=&\frac{e^{-\frac{1}{2}\gamma s}}{2\Omega_1 \Omega_2}\left[\Omega_2\sin(\Omega_1 s)-\Omega_1\sin(\Omega_2 s)\right]
\\
v_{--}(s)&=&\frac{e^{-\frac{1}{2}\gamma s}}{4\Omega_1 \Omega_2(\Omega_2^2-\Omega_1^2)}\left[\Omega_2\sin(\Omega_1 s)\left(\gamma^2-8\gamma\Omega_1-2\Omega_1^2+2\omega_2^2\right)\right. \nonumber \\
&& \left.+ 4\gamma\Omega_1\Omega_2\left(\cos(\Omega_2 s)-\cos(\Omega_1 s)\right)-\Omega_1\sin(\Omega_2 s)\left(\gamma^2-8\gamma\Omega_1+2\Omega_1^2- 2\Omega_2^2\right)\right] \hspace{1cm}
\end{eqnarray}
From $v(s)$ one constructs the matrix $R$ using Eq. (\ref{ceq2}) and the matrix $S$ using Eqs. (\ref{sxx}---\ref{sxp}). To obtain the asymptotic state, we compute $S$ at the limit $t \rightarrow \infty$. The off-diagonal elements in the $X_+, X_-$ basis vanish, while
\begin{eqnarray}
S_{X_+X_+}&=&\frac{\gamma}{M \pi}\int_0^\infty d\omega \omega f(\omega) (-2\omega^2+\Omega_1^2+\Omega_2^2)^2 ,
\\
S_{P_+P_+}&=&\frac{4M\gamma}{\pi}\int_0^\infty   d\omega \omega^3 f(\omega) (-2\omega^2+\Omega_1^2+\Omega_2^2)^2 ,
\\
S_{X_-X_-}&=&\frac{\gamma}{M\pi} (\Omega_1^2-\Omega_2^2)^2\int_0^\infty d\omega  \omega , f(\omega)
\\
S_{P_- P_-}&=&\frac{4M\gamma}{\pi}(\Omega_1^2-\Omega_2^2)^2 \int_0^\infty d\omega  \omega^3 f(\omega),
\end{eqnarray}
where
\begin{eqnarray}
f(\omega) = \frac{e^{-\frac{\omega^2}{\Lambda^2}}\coth\left(\frac{\omega}{2T}\right)}
{[2(\omega^2-\Omega_1^2)^2+\gamma^2(\omega^2+\Omega_1^2)]
[2(\omega^2-\Omega_2^2)^2+\gamma^2(\omega^2+\Omega_2^2)]}.
\end{eqnarray}
The asymptotic values for $S$ can be evaluated to leading order in $\gamma/\Omega_i$ and $\Omega_i/\Lambda$, by substituting the Lorentzians in the integrals with a delta function, that is, $[(x-a)^2+\gamma^2]^{-1} \simeq \frac{\pi}{2 \gamma} \delta (x-a)$. This corresponds to the weak-damping limit of Ref. \cite{HZ}. We then obtain
\begin{eqnarray}
S_{X_+X_+} = S_{X_-X_-} = \frac{1}{8M} \left( \frac{\coth\frac{\Omega_1}{2T}}{\Omega_1} + \frac{\coth\frac{\Omega_2}{2T}}{\Omega_2} \right), \\
S_{P_+P_+} = S_{P_-P_-} = \frac{M}{2} \left(\Omega_1 \coth\frac{\Omega_1}{2T} + \Omega_2 \coth\frac{\Omega_2}{2T} \right),
\end{eqnarray}
which correspond to an asymptotic thermal state for the pair of oscillators.

\end{appendix}

\end{document}